\documentstyle[isolatin1,twocolumn,amssymb,aps,prb]{revtex}

\def\markthis{}

\begin{document}

\wideabs{
\title{Crater formation: can macroscopic scaling laws be used in microscopic
cratering?}
\author{E. M. Bringa$^1$, K. Nordlund$^2$ and J. Keinonen$^2$}

\address{$^1$ Engineering Physics and Astronomy Department. University
of Virginia. Charlottesville VA 22903 U.S.A.}

\address{$^2$Accelerator Laboratory, P.O. Box 43, FIN-00014 University of 
Helsinki, Finland}

\date{\today}
\maketitle

\begin{abstract}
Using classical molecular dynamics simulations we examine the formation of
craters during 0.4 - 100 keV Xe bombardment of Au. Our simulation results,
and comparison with experiments and simulations of other groups, are used to
examine to what extent analytical models can be used to predict the size and
properties of craters. We do not obtain a fully predictive analytical model
(with no fitting parameters) for the cratering probability, because of the
difficulty in predicting the probability of cascades splitting into
subcascades, and the relation of the heat spike lifetime and energy density.
We do, however, demonstrate that the dependence of the crater size on the
incident ion energy can be well understood qualitatively in terms of the
lifetime of the heat spike and the cohesive energy of the material. We also
show that a simple energy density criterion can not be used to predict
cratering in a wide ion energy range because of the important role of the
heat spike lifetime in high-energy cascades. The cohesive energy dependence
differs from that obtained for macroscopic cratering (observed e.g. in
astrophysics) because of the crucial role of melting in the development of
heat spikes.
\end{abstract}


}

\sloppy
\flushbottom

\thispagestyle{myheadings} 
\markboth{}{Preprint, submitted for publication to Phys. Rev. B (2001)
\quad }

\section{Introduction}

Surface modification of materials by incident ions has been observed using
electron microscopy, scanning tunneling microscopy (STM) and atomic forces
microscopy (AFM) \cite{Jag88,papaleo,papaleo0,balanzat}. A large variety of
features have been studied:\ hillocks \cite{bumps}, depressions \cite
{Ade00,Reimann-crat,reimann-crater,vorobyova}, crater rims 
\cite{Jag88,papaleo,papaleo0,Bir96}, 
adatoms \cite{Nor98c,Kyu99,Gha97,Gha94}, and surface roughening \cite
{aoki-thesis}. Hillocks can appear when an energetic process occurs a few
layers below the surface. For instance, an energetic recoil can create a
mini-spike which melts the surrounding region, creating a low density region
of larger volume which raises the surface \cite{Ave98}. When the energy
loss per unit path length of the projectile, $dE/dx$, and\ the sputtering
yield are relatively small, adatoms are observed in both experiments and
simulations. For larger energy deposition (and larger yields) a crater is
formed. For even larger yields re-deposition of the ejecta plus plastic
deformation occurs producing craters with rims, studied recently for ion
bombardment of polymers \cite{papaleo}. Craters are also produced by cluster
ion bombardment in which non-linear effects lead to enhanced sputtering \cite
{andersen,yamada1,yamada2,yamada3,Colla2000}. This has been studied in the velocity
regime in which nuclear (elastic) energy loss dominates over electronic
energy loss and has also been observed in experiments \cite
{andersen,yamada1,yamada2,yamada3}.

Although several studies have used molecular dynamics (MD) computer
simulations to examine crater formation 
\cite{Gha94,Bri98,Ade00,shulga-clusters,averback-clusters,Colla2000,%
averback-clusters1,insepov-nato,insepov,insepov1}, most of them
have been limited to fairly low total ion or cluster impact energies, $%
\lesssim $ 10 keV. In this energy range, a fairly good description of the
mechanisms of cratering has been found in dense fcc metals \cite
{Ade00,insepovPRB}. Yet MD simulations of mixing in the bulk have shown that
the amount of atom displacements keeps increasing superlinearly up to ion
energies of $\sim $ 100 keV in dense fcc metals, and that the heat spikes
formed can persist for tens of picoseconds \cite{Nor98}. This poses the
question whether the long lifetime of high-energy spikes can alter the
mechanism of cratering in fcc metals, as significant liquid flow might take
place in cascades with long lifetimes. In fact, in a recent paper Aderjan
and Urbassek \cite{Ade00} suggest that liquid flow might explain why the
cratering cohesive energy dependencies differ from those predicted by
macroscopic models \cite{Gault75,lampson,quinones98,explosion-cratering}, but
they do not give a description of the mechanism by which this might occur.

In this paper, we use MD simulations to examine crater formation by Xe
recoils impacting on Au in an unprecedentedly wide energy range, ranging
from 0.4 up to 100 keV of initial ion energy. We chose this ion-solid
combination because TEM experiments are available for cratering in the same
system \cite{Bir96,Don97,Bir2000}. We find that at higher energies the fundamental
mechanism leading to cratering does indeed change. While in the low-energy
regime ($\lesssim 10$ keV) the mechanism can be understood on the basis of
a high kinetic energy density alone, in agreement with previous models for
dense metals, we show that at high energies ($\gtrsim 50$ keV) cratering
can result from lower kinetic energy densities due to the long lifetime of
the heat spike. We also present analytical models for both energy regimes,
and compare our simulation results with experiments and results in other
types of materials.

This paper is organized as follows. First some details of the simulation and
crater identification and measurement are given in section \ref{method}. In
section \ref{individual} results for a few individual events are presented,
leading to the identification of weak points in current analytical models of
cratering. In the following section we discuss the formation mechanisms of
our craters in detail, and relate them to previous models. In section \ref
{expt_comp} we compare our results directly with experiments, and in section 
\ref{binding_role} discuss why macroscopic scaling laws do not apply to
atomic systems. We then compare our results for single-ion bombardment with
results for cluster bombardment in section \ref{cluster_comp} and finally in
section \ref{othermater} show that with appropriate scaling crater sizes in
both metals and some organic solids can be understood in the same framework.

\section{Method}

\label{method}

\subsection{Molecular dynamics simulations}

The basic MD simulation methods used in this work has been described in
several previous papers \cite{Gha97,Nor97f,Nor98,Nature}, so in here we only
recall the basic principles, and the features which differ from those. In
simulating ion irradiation of a surface, we place an incident ion on a
random position a few \AA {} above the surface, and give it a kinetic energy
of 0.4~--~100 keV towards the sample. The incident angle is chosen in an
off-channeling direction close to the surface normal. Most simulations were
carried out for a (001) surface, but six 100 keV events were also simulated
for a (111) surface. No major difference were observed in the crater sizes
for the two surfaces for 100 keV energies. The development of the system of
atoms is followed until the cascade has cooled down close to the ambient
temperature, which for the cascades presented here was always 0 K to
definitely rule out any post-cascade damage annealing. A few 50 keV events
(the results of which are not presented here) were also simulated at 300 K,
and found to give similar crater sizes as the 0 K events. Varying the
initial position of the recoil atom can cause the resulting cascades to
behave very differently depending on where the strongest collisions occur.

The simulation cells had periodic boundaries in the $x$ and $y$ dimensions,
and a fixed bottom layer in the $z$ direction. They were cooled down to 0 K
using Berendsen temperature control at the cell borders and a few layers at
the bottom above the fixed layer. The simulation cells had at least 16 atoms
per eV of incident ion energy, which was enough to prevent cell heating
beyond the melting point or pressure wave reflection from the borders strong
enough to affect the cascade outcome. To ensure that no artificial border
effects occurred, runs were automatically stopped and restarted in a larger
simulation cell if a recoiling atom with an energy higher than 20 eV (which
is less than the threshold displacement energy \cite{And79}) entered the
temperature scaling region.

Most simulations were carried out in Au modeled by the embedded atom method
(EAM) potential of Foiles \cite{Foi86}, smoothly joined to the universal
repulsive interatomic potential of Ziegler, Biersack and Littmark \cite{ZBL}
at small interatomic separations \cite{Gha94}. To probe the cohesive energy
dependence of the crater size, we used artificial modifications of this
potential for different values of the cohesive energy. To be precise, in the
EAM formalism \cite{Foi86,Daw93} we scaled the pair potential and the
embedding energy of the potential by a factor $f$ in the range 0.5~--~1.6,
while leaving the electron density unmodified. To preserve the ballistic
properties of the potential, however, the factor $f$ was scaled smoothly
back to 1.0 at small interatomic separations $r$, in the same $r$ range
where the repulsive potential was onset. This scaling gives an interatomic
potential with the same equilibrium lattice constant as the original
potential, but with the cohesive energy $U_0$ and elastic moduli scaled by
the factor $f$. We further verified by simulations that the melting point $%
T_{{\rm melt}}$ scaled highly accurately by $1/f$, as expected from
Lindemann's law \cite{Lin10}.

For reference purposes, we note that at 0 K the binding energy of our EAM Au
is $U_{0}=3.930$ eV, the lattice constant $a=4.080$, and hence the atomic
density is $n=0.0589$ \AA $^{-3}$. The melting temperature of the simulated
solid is 1110$\pm 20\,$K.

\subsection{Finding non-channeling directions}

To enable the use of simple binary collision approximation programs such as
TRIM \cite{ZBL} for quick estimates of cascade energy densities and
penetration depths, it is important to use an incident angle in the
simulation corresponding to a non-channeling direction. To find such a
direction, we used the MDRANGE code, which is an ion range calculation code
which accounts for the crystal structure \cite{Nor94b}.

\vbox{
\begin{table}
\caption{
Simulated mean range ($\bar R$) and straggle $S_R$ values 
for 50 keV Xe bombardment of (001) Au surfaces at 0 K,
using no zero-point atom displacements.
Using thermal atom displacements corresponding to
300 K gave almost as strong channeling effects,
for instance giving a 
mean range $\bar R = 103 \pm 3$ \AA\ for $\theta= 10^\circ$, 
$\phi=20-30^\circ$.
The ion range is here defined as the ion penetration depth.
$\theta$ is the angle between the initial ion
velocity vector and the (001) surface normal (``tilt'' angle), 
and $\phi$ the angle between the initial vector and the (100)
direction in the surface plane (``twist'' angle).
All angles are given in degrees. 
A range of angles such as ``0-360'' means the angle was selected
randomly in this range.
}
\label{rangetable}
\begin{tabular}{llcc}
$\theta$	& $\phi$	& $\bar R$ (\AA)& $S_R$ (\AA) 	\\
\hline
0               & 0             & $1240\pm50$       & 780   \\
5               & 20-30         & $670\pm20$        & 480   \\
10              & 20-30         & $108\pm5$         & 110   \\
15              & 20-30         & $81\pm3$          & 90    \\
20              & 20-30         & $79\pm3$          & 60    \\
25              & 20-30         & $69\pm3$          & 60    \\
30              & 20-30         & $75\pm2$          & 76    \\
25              & 0-10          & $164\pm14$        & 251   \\
25              & 10-20         & $73\pm2$          & 70    \\
25              & 20-30         & $69\pm3$          & 60    \\
25              & 30-40         & $74\pm3$          & 60    \\
\end{tabular}
\end{table}
}

\pagebreak
\vbox{
\widetext
\begin{figure}
\includegraphics{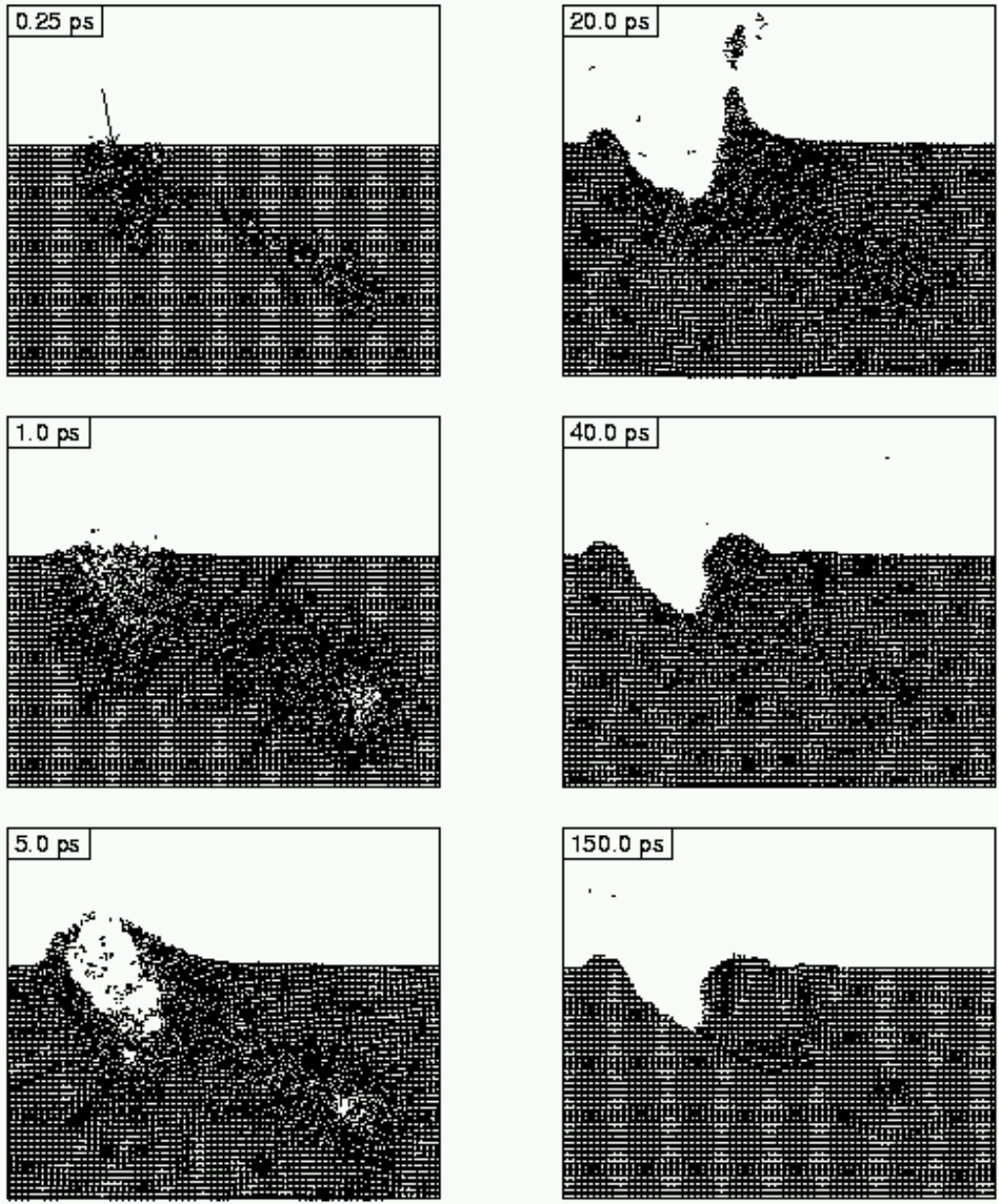}
\nopagebreak \vskip 530pt minus 0pt \nopagebreak
\caption{
(({\it lanl preprint n.b.: figure will be in
better resolution in final published paper}))
Crater formation by a 100 keV Xe ion hitting a (001) Au surface.
The figure shows a cross-sectional slice of 8 atom layers in the $(\bar 2 1
0)$ plane in the central part of the simulation cell.
This cross-section was chosen to give a good illustration of the subcascade
splitting. The arrow in the first frame indicates the initial direction of
the incoming ion. The tip of the arrow shows the impact point of the ion
projected on this cross section; the actual impact did not occur in these
atom layers. The 0.25 ps and 1.0 ps snapshots show that the cascade splits
into two almost separated subcascades during the ballistic phase of the
cascade. The subcascade below the surface subsequently behaves like a
cascade in the bulk \protect\cite{Nor97f}. 
The subcascade at the surface produces a
crater between 2 and 40 ps, and also causes a large atom cluster to 
sputter \protect\cite{Bir2000}.
This sputtered cluster is so hot that it emits a large number of Au atoms
and dimers, many of which redeposit on the surface. Notice also how an
interstitial-like dislocation loop has formed at 20 ps close to the right
edge of the crater, and at 40 ps has produced an adatom island next to the
crater by coherent displacement \protect\cite{Nor98c}. The final 150 ps 
snapshot shows the final crater structure, a vacancy loop produced by the 
subcascade inside the sample, and a complex dislocation structure on the 
right-hand side of the crater.
}
\label{snapshots}
\end{figure}
\narrowtext
}
\pagebreak

\phantom{a}

\pagebreak

\vbox{
\begin{figure}
\includegraphics{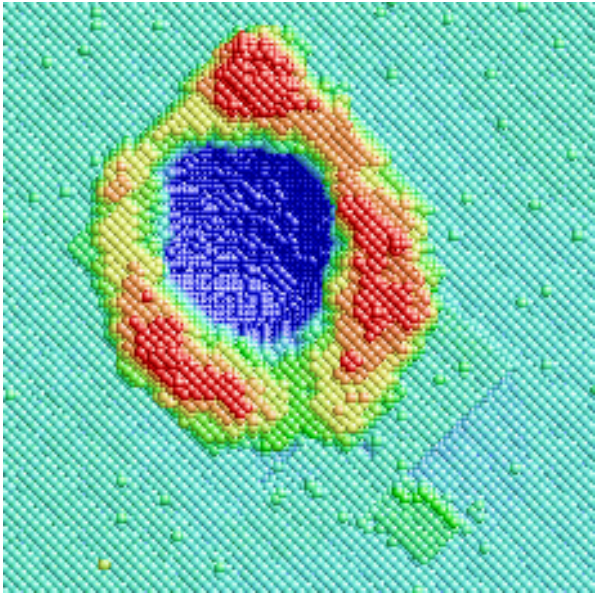}
\nopagebreak \vskip 8cm minus 0pt \nopagebreak
\caption{ (({\it lanl preprint n.b.: figure will be in
better resolution in final published paper}))
(color)
Crater produced in the event illustrated in Fig. \ref{snapshots}
seen from above. The colors indicate the height of the atoms. Blue and cyan
atoms are below the original surface, and green, yellow and red atoms above,
with the red atoms being highest up. The color scale has been chosen to
emphasize atom layers at the surface; hence all atoms deeper than 1 unit
cell (4.08 \AA{}) below the original surface have the same blue color. The
maximum depth of the crater is about 44 \AA{}. Notice the regular adatom
island (green atoms) on the lower right side of the crater, produced by the
coherent displacement mechanism. The dislocation below the surface also
produces a regular atom edge close to the crater, just next to the adatom
island. The single adatoms far from the crater are atoms which have been
redeposited on the surface from the sputtered atom clusters.
}
\label{crater}
\end{figure}
\narrowtext
}

In the particular case of heavy ion irradiation of (001) fcc metals, we have
observed that channeling effects can be extremely strong, requiring careful
selection of both the tilt ($\theta $) and twist ($\phi $) angles to obtain
a good non-channeling direction. As we shall see below, this may be of
crucial importance for the comparison of cratering probabilities with
experiments, whence we give some sample range results for 50 keV Xe
bombardment of (001) Au at 0 K in Table \ref{rangetable}. The results show
that the mean range $\bar{R}$ depends sensitively on both $\theta $ and $%
\phi $. Ranges at other energies had a quite similar $\theta $ and $\phi $
dependence. To obtain an optimal non-channeling direction we used $\theta
=\phi =25^{\circ }$ for the (001) surface at all energies.

\subsection{BCA Calculations}

A number of SRIM 2000.39 \cite{SRIM2000} calculations were performed in
order to compare binary collision approximation (BCA) results with MD
results. If good agreement is found e.g. in the energy densities, the SRIM
code could be used to obtain quick estimates of cratering probabilities. The
runs used the same incident angle, $\theta =25^{\circ }$, $U_{surf}$=2.6 eV,
and $U_{bulk}$=3.9 eV.  We used $E_{D}=25$ eV as the displacement
energy value for SRIM \cite{And79}. 
The low energy self-sputtering of Au at normal incidence was
reproduced well with these values of the binding energies and displacement
energy. The total energy deposited up to certain depth, $E_{kin,tot}$ can be
calculated by integrating the energy deposited as a function of depth $z$, $%
F_{D}(z)$. Energy densities were calculated using a cylindrical volume
(which contained the cascade better than a hemispherical volume). SRIM
follows recoils until they reach the displacement energy, and therefore one
should really compare with the MD energy density discarding atoms with $%
E_{kin}>E_{D}$. However, the energy deposition in SRIM represents the
scenario before 0.1 ps, when there are only few energetic recoils and the
cascade has not evolved significantly.

\subsection{Analysis}
\label{MD-Analysis}

Post-run analysis of the quenched MD simulation cells was performed to
identify the surface features. If the last layer of atoms was located at $%
z=0 $, with the target in $z<0$, all atoms in the range $a/4<z<r_{sput}$
were counted as adatoms, and all atoms above $r_{sput}$ were counted as
sputtered atoms. $r_{sput}$ was set to 20 \AA\ for all cases except the 100
keV ones, where $r_{sput}=40$ \AA. Both values are much larger than the
cut-off radius of the potential, which is only 5.55 \AA .

The crater and the crater rims were measured along two directions and the
mean sizes were evaluated. The deviation of the crater shape from a circle
was estimated with a parameter $\alpha =R_{<}/R_{>}$, where $R_{<}$ and $%
R_{>}$ are the smaller/larger crater radii, respectively. This parameter was
typically 0.6~--~0.9, since the craters were not circular but diamond
shaped, following the symmetry of the lattice, as seen in cluster
bombardment simulation of Cu \cite{Ade00}. The area of the craters was
therefore calculated as a circle and also as a parallelogram, which gives
smaller areas, but within 10~--~30\% of the value for a circular area.

The rims were identified as structures with more than one atomic layer above
the surface, to differentiate them from single-layer adatom islands and
coherent displacement \cite{Nature}, which occurs often at the side of the
rim for the 20~--~100 keV events.

The volume of the crater is approximated as the number of atoms `excavated'
from the target, $nV=N_{adat}+N_{sput}=N_{tot}$, where $n$ is the
equilibrium atomic density of the solid. If the number of atoms on top of
the surface is used as an estimate of the crater volume, there are two small
corrections which have not been taken into account. There may be atoms far
from the crater which are on top of the surface because of coherent
displacements. Besides, there may be interstitials or vacancies in the
crater walls which change the normal density of the material, but their
number was observed to be small, and in any case their effect seems to
roughly cancel each other out. The `measured' radius of the crater is not
affected by these corrections.

\pagebreak
\widetext
\vbox{
\def\underbar#1{#1}
\begin{table}
\caption{
Results for individual cascade events simulated by MD and SRIM
averages for the same energies. For MD, $E_{kin,tot}$ is the total energy of 
the liquid atoms in the top 40 A, averaged over two times around 0.1 ps (for
instance at 0.07 ps and 0.15 ps). $E_{kin,mean}$ is the mean energy of
all atoms inside a cylinder of radius $R_{rad}$ and length Depth.
For the MD simulations $R_{lat}$, $R_{rad}$ and ``Depth'' 
indicate the size of the
cascade. $\varepsilon $ is the energy density calculated 
using a cylinder containing the cascade, calculated
at 0.1 ps in the MD runs, 
$n$ is the atomic density and $U_0$ the equilibrium
potential energy of the material.
$N_{vac}$ is the number of vacancies, and $Y$ is the
initial sputtering yield. Note that because many of the sputtered
atoms leave the surface in hot clusters, and can subsequently evaporate from
the cluster and redeposit on the surface, the initial MD sputtering yield 
is not exactly comparable to experimentally measured yields.
Individual simulations are labeled by an extra number, for
instance, case number 8 for 10 keV bombardment is labeled 10-8. 
}
\label{table:individual}
\begin{tabular}{|c|c|c|c|c|c|c|c|c|c|}
	& \multicolumn{5}{c|}{\underbar{5 keV}} & \multicolumn{2}{c|}{\underbar{10 keV}} 
      & \multicolumn{2}{c|}{\underbar{100 keV}} \\ \hline 
Event	& 5 & 5-2 & 5-1 & 5-4 & 5-9 & 10 & 10-8 & 100 & 100-4\\ \hline
Model & SRIM & MD & MD & MD & MD & SRIM & MD & SRIM & MD \\ \hline
Crater? & --- & no & small & yes & yes & --- & yes & --- & yes \\ \hline
$E_{kin,tot}$ (keV) & 4.42 & 2.125 & 2.81 & 2.71 & 3.16 & 8.32 & 6.27 
& 25.4 & 62.35 \\ \hline
$E_{kin,mean}\,$(eV) & 1.84 & 0.88 & 1.17 & 1.13 & 1.31 & 1.66 & 2.4 
& 0.31 & 0.76 \\ \hline
$R_{crat}$ & --- & --- & 8$\pm 3$ & 16$\pm 4 $ & 16$\pm 4$ & --- & 
19.6$\pm 2$ & --- & 36$\pm 6$ \\ \hline
$R_{lat}$/$R_{rad}$ & 13/18 & 20 & 25 & 17 & 12.5 & 19/26 & 20 & 77/105 & 60
\\ \hline
Depth & 30 & 40 & 30 & 40 & 25 & 40 & 35 & 250 & 60 \\ \hline
$\varepsilon /nU_0$ & 0.8 & 0.39 & 0.43& 0.68 & 2.35 & 0.5 & 0.62 & 0.07& 0.21 \\ 
\hline
$N_{vac}$ & 119.5 & 75 & 73 & 299 & 152 & 233.3 & 408 & 1,064.5 & 4,538 \\ 
\hline
$Y$ & 15.9 & 2 & 6 & 31 & 15 & 21.3 & 44 & 36.3 & 472 \\ 
\end{tabular}
\end{table}
}

\vbox{
\begin{table}
\caption{
Summary of simulation results for different energies. To calculate 
outflow time using Eq. \ref{tflow0}, and spike times in Eq. \ref{tspike}, 
we used $R_c=R_l$ for 1-20 keV, 
$R_c = 0.75 R_l$ for 50 keV, $R_c = 0.50 R_{l}$ 
for 100 keV. The spike times for the bulk of the 
MD simulation were calculated as the time when the number of liquid atoms 
was reduced by a factor of 10 from its maximum value, which occurs at 
$\approx$ 0.5 ps. The spike times for the surface were calculated as the 
times when the number of liquid atoms at the surface had a maximum. 
Cratering probabilities were evaluated as explained in the text.}
\label{table:averages}
\begin{tabular}{|c|c|c|c|c|c|c|}
 $E_0$ (keV) & 1 & 5 & 10 & 20 & 50 & 100 \\ \hline
events & 20 & 10 & 20 & 20 & 10 & 6 \\ \hline  
$R_{c}$ (\AA )(MD) & 8.5$\pm 1.5$ & 13$\pm 4$ & 19.6$\pm 2$ & 24$\pm $4 & 31%
$\pm $2.5 & 33$\pm 2.5$ \\ \hline 
$R_{l}$ (\AA )(SRIM) & 7 & 13 & 19 & 26 & 43 & 65 \\ \hline 
$t_{spike}$ (ps)(MD)-bulk & 3.8 $\pm 0.2$ & 9.7 $\pm 0.4$ & 13 $\pm 3$ & 13 $\pm 1$ & 33$\pm 3$ & 30$\pm 10$ \\ \hline 
$t_{spike}$ (ps)(MD-surf)-no crater & 0.9 $\pm 0.3$ & 2.3 $\pm 0.4$ & 5 $\pm 2$ & 18 $\pm 2$ & 7.5 $\pm 3$ & 4$\pm 2$ \\ \hline 
$t_{spike}$ (ps)(MD-surf)-crater & 0.9 $\pm 0.2$ & 2.6 $\pm 0.4$ & 6 $\pm 2$ & 23 $\pm 1$ & 12 $\pm 3$ & 16$\pm 5$ \\ \hline 
$t_{spike}$ (ps)(Eq. \ref{tspike}) & 0.8 & 2.8 & 6.0 & 11.3 & 17.3 & 17.6 \\ \hline 
$t_{outf}$ (ps)(Eq. \ref{tflow0}) & 2.6 & 4.9 & 7.1 & 9.8 & 21.0 & 21.2 \\ \hline 
$P\left[ \varepsilon \left( E_{o}\right) >\varepsilon _{c}\right] $ (MD) &0.80 & 0.78 &0.73 &0.7& 1 &1 \\ \hline 
$P_{split}$ & 0.0 & 0.1 & 0.2 & 0.3 & 0.3 & 0.4 \\ \hline
$P\left( E_{o}\right) $ (MD) & 0.79 & 0.67 & 0.60 & 0.55 & 0.83 & 0.60 \\ \hline 
$P\left( E_{o}\right) $ (Eq. \ref{Pcratering}) &  0.80 & 0.70 & 0.58 & 0.49 & 0.70 & 0.60 \\
\end{tabular}
\end{table}
\narrowtext
}
\pagebreak

\phantom{a}

\pagebreak

\section{Individual cascades and cratering}

\label{individual}

A detailed picture of the cratering formation scenario can be obtained by
considering individual MD simulations. For brevity, we use the following
notation to denote individual events. The B'th event, were B is a running
index, carried out with an energy of ``A'' keV is denoted by ``A-B''. Thus
for instance the eighth 10 keV event is marked by ``10-8''. Results for a
few events are listed in Table \ref{table:individual} and discussed below.

Fig. \ref{snapshots} shows cross-sectional snapshots of a 100 keV cascade,
event 100-4, with a crater being formed between 2 and 40 ps. Fig. \ref
{crater} shows the final crater produced in this event. The liquid flow of
atoms builds the crater and crater rim. Notice that the cascade splits below
the surface, and one of the subcascades does not reach the surface, even
though it produces some coherent displacement \cite{Nor98c} of atoms next to
the crater. Below we analyze in detail a few cascades in order to clarify
some results regarding cratering probability, which is discussed in the
following section.

In Table \ref{table:individual} we compare MD results for individual events
to average results from the SRIM 2000 BCA code. The MD values for cascade
size, mean energy per atom, and so on compare reasonably well with SRIM,
except for 100 keV. Note that the energy density at 0.1 ps, when the craters
just start to form, is of the order of 0.3~--~2$nU_{0}$. For the 10 keV case
shown in the table, by 1 ps this energy density decreased to $\sim
0.075nU_{0}$, in good agreement with other estimates of mean kinetic energy (%
$E_{kin}$) in the molten region of the cascade \cite{Ave94b,bumps}. The case
10-8 forms an almost hemispherical crater, as well as the case 5-9. The case
5-4 has a large subcascade splitting, but both cascades are very close to
the surface. On the other hand, case 5-2 has two subcascades, with some
coherent atomic displacement on top of the larger cascade producing only few
adatoms and a platelet. The case 100-4 forms a large crater, but notice that
the energy density is lower than that for the 5 keV event that did not
produce any crater. Since the cascade is so large, it can stay hot longer,
and flow occurs during tens of ps, even though the energy density is low.
Notice that the crater radius is much smaller than the cascade size, but the
rim for 100-4 has a length of $100\pm 10$ \AA , close to the lateral range
of 105 \AA .

From the analysis of the above cases and other events not discussed in 
detail here, we can already qualitatively conclude that:

\begin{enumerate}
\item  At low bombarding energies, cratering occurs for energy densities
close to 0.5$nU_{0}$.

\item  Energy density alone is not a good criterion for crater formation,
since for large spikes the long lifetime of the spike will allow ejection
even for energy densities much lower than $nU_{0}$. To obtain a cratering
probability one would need to take into account the probability of the spike
being long lived, related to both spike radius and available energy.

\item  Cascade splitting should somehow be included, because it can dilute
the energy density significantly and concentrate it away from the surface.
As a result the cratering probability is reduced.
\end{enumerate}

\section{Formation mechanisms}

\label{mechanisms}

Averaged simulation results can be found in Table \ref{table:averages}. The
data for $R_c$ includes only the results of those simulations which 
produced a crater. The error bars indicate standard deviation of those 
events. The cratering probability was calculated as the ratio of cratering 
events to total number of events, $P\left( E_{o}\right) =N_{crat}/N_{total}$, 
and it was always larger than 40\% for the energies studied.

\subsection{Role of spikes at different energies}

In the early work of Thompson and Johar \cite{Tho79}, where large deviations
from linear cascade theory were found, it was proposed that thermal spikes
were not necessary, and that a decrease in the surface binding would be
enough to explain the data. In our simulations spikes are certainly playing
a large role and, since the surface damage is considerable, the surface
binding also decreases.

It has been claimed that the energy density in the cascade determines the
crater formation \cite{Jag88,Bir96,Don97}, 
and that an energy density larger than 
$nU_{0}$ is needed to produce a crater. On the other hand, several models 
\cite{Ave94b,bumps} assume that the flow of liquid atoms towards the surface
can occur when atoms have $E_{kin,mean}=3k_{B}T_{m}=0.29\,$eV, giving a much
lower energy density than before, $\varepsilon =0.07nU_{0}$. We now use our
simulation results to elucidate the reasons behind this apparent discrepancy.

In order to verify the first assumption, a simple estimate can be made as
follows. For instance, for $E_{o}=50$ keV bombardment, we can assume an
hemispherical crater of radius $R_{c}$, and that $E_{o}$ is shared by all
atoms inside the crater, which gain an energy $E_{c}$, $\left( 2/3\right)
\pi R_{c}^{3}nE_{c}=E_{o}$. Using $R_{c}=30$ \AA , close to the one found in
MD (see Table \ref{table:averages}), gives $N_{crat}=$3336, $E_{c}=15\,$%
eV/atom $\sim 3.8 U_{0}/$atom, if most of the energy is originally deposited
inside the crater region. $N_{adatoms} \approx 3370$ from MD compares well
with the missing crater volume. The agreement is quite good, except for
energies below 20 keV, when craters are shallow, or energies above 50 keV,
where a significant fraction of the energy is deposited deep inside the
sample. However, as can be seen from Table \ref{table:individual}, the
energy density at energies above 10 keV is much smaller than $nU_{0}$,
although higher than $0.07nU_{0}$. At lower energies the energy density
increases, but still is smaller than the minimum energy that would produce
craters according to Ref. \onlinecite{Jag88}. The cascade in the simulation
10-8, in Table \ref{table:individual}, has a crater size $R_{crat}=19.6\pm 2$
\AA , and\ $E_{kin,tot} = 6.27$~keV. Inside a cylinder of radius 20 \AA ,
and height 35 \AA\ there are 2595 atoms, and $E_{kin,mean}=2.4$ eV $< U_{0}$%
. This gives $\varepsilon =0.62\,nU_{0}$ ($nU_{0}=0.2315$ eV/\AA $^{3}$).
Discarding the atoms with kinetic energy below $U_{0}$ gives $\varepsilon
_{c}\sim 0.59nU_{0}$.

Thus we see that the apparent contradiction in energy densities mentioned
earlier may at least in part arise from the fact that the energy density
needed for crater production does in fact strongly depend on the spike
lifetime.

Merkle and J\"ager proposed that surface spikes were the cause of the
cratering and sputtering yield enhancement they observed in Bi$^{+}$ $\,$and
Bi$^{++}$ bombardments of Au. Crater formation in our simulations is related
to the probability of the cascade being close to the surface (see Section 
\ref{craterprob}), as in their work. However, since they could only see
craters with a radius larger than 25 \AA {}, they find a high energy
threshold for crater formation. In our simulations we observe that crater
formation can still occur at relatively low bombarding energies, but the
crater size is well below the detection limit of Merkle and J\"ager (see
Fig. \ref{crmean}). They also assume that atoms originally resident at the
crater site are all sputtered, while we now know that most of the atoms will
be redeposited at the crater rim and will not contribute to the sputtering
yield.

Based on the above discussion and the results in Section \ref{individual},
we can identify two regimes in the crater formation process. First, at low
energy deposition only a relatively small hot region is created. For single
ions this will occur when the penetration depth has roughly the same size as
the lateral range of the ion, i.e. up to 10~--~20 keV in Xe $\rightarrow $
Au. If the energy density deposited in the cascade is larger than $%
0.25nU_{0} $ a crater will be formed, provided the cascade is connected to
the surface and did not decay into multiple subcascades. The probability of
crater formation is slightly lower than one only because of the latter
reason. For energies lower than 1 keV, the lateral size of the cascade is of
the order of the lattice spacing, and it would be difficult to distinguish a
crater from a small vacancy cluster. We did perform a number of 400 eV
events, and found that 4 out of 17 events simulated resembled craters.
However, because of the difficulty of clearly defining what is a crater with
the very small number of atoms involved, we did not include these events in
the quantitative analysis. In the low-energy (1~--~20~keV) regime, the
crater has a roughly hemispherical shape, the radius of the crater is close
to the radius of the lateral range of the ion in the solid, and the
dependence of the size with energy follows the $E_{o}^{1/3}$ law. Prompt
sputtering of hot atoms cools down the spike significantly.

Second, at high energy deposition, the spike region is large, and the energy
density is lower than $0.25nU_0$. However, since the spike is large, the
center of the spike cools down much slower that its sides, and there are
atoms which can flow out creating a crater. The crater radius is only a
fraction of the lateral size of the cascade, since the borders cool down
rapidly. Writing

\begin{equation}
R_{c}\sim R_{l}\left[ 1.13-2.6\ 10^{-5}(E_{o}/U)\right] ,  \label{rc-rl}
\end{equation}

gives a very good estimate of the crater radius as a function of the lateral
size of the cascade.

Next we discuss this liquid flow in greater detail.

\vbox{
\begin{figure}[tbh]
\includegraphics{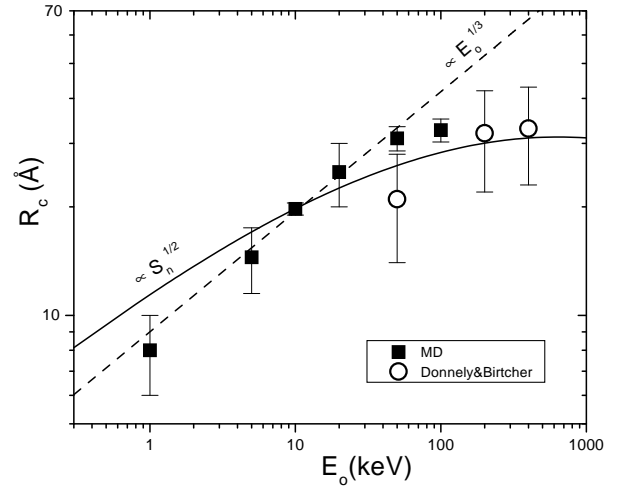}
\nopagebreak \vskip 200pt minus 0pt \nopagebreak
\caption{ 
Crater radius $R_c$ as a function of the incident energy $E_{o}$,
including experimental data of Donnelly and Birtcher \cite{Bir96,Don97}.
}
\label{crmean}
\end{figure}
}

\subsection{Liquid Flow}

There will be a crater if the time for atomic outflow, $t_{outf}$, is close 
to or larger than the lifetime of the spike, $t_{spike}$. $t_{outf}$ can be 
estimated as
the time it takes the atoms originally inside the crater to flow out.
Assuming for simplicity a cylindrical outflow at constant velocity $v_{c}=%
\sqrt{2E_{c}/m}$, and up to a depth $R_{c}$,

\begin{equation}
n\frac{v_{c}}{4}\pi R_{c}^{2}t_{outf}=n\pi R_{c}^{3}\Rightarrow t_{outf}=%
\frac{4R_{c}}{v_{c}}  \label{tflow0}
\end{equation}
$v_{c}$ is a mean velocity of the atoms after the outflow begins. Writing $%
E_{c}=\alpha U_0$, $t_{outf}=%
{\displaystyle {4R_{c} \over \sqrt{\alpha }\overline{v_{u}}}}%
=0.84\left( R_{c}/a\right) /\sqrt{\alpha }$. 
(Ref. \onlinecite{oldbarriernote}). 
The spike lifetime can be estimated assuming lattice heat conductivity
with a constant heat diffusivity $\kappa$ (Ref. \onlinecite{noelphonon}).
Then, 
\begin{equation}
t_{spike\approx }\sim R_{c}^{2}/\left( 4\kappa \right) .  \label{tspike}
\end{equation}

If the lifetime of the spike is estimated as the time that it takes to cool
down from the initial energy density to the critical energy density, $%
t_{spike}$ can also be obtained analytically (assuming a Gaussian
temperature profile), given the initial cascade parameters, and gives times
within 30\% of the simple estimate in Eq. \ref{tspike}. In order to check
the validity of Eqs. \ref{tflow0} and \ref{tspike}, 
we compared the times from those
equations to the MD results, using $\kappa =15$ \AA $^{2}$/ps (obtained from
MD), and Eq. \ref{rc-rl} to obtain the spike radius. Note that typically it
is assumed that $r_{spike}^{2}\propto E_{o}$ (Ref. \onlinecite{Ave94b}) but,
following transport theory, the lateral range is roughly linear with $E_{o}$
which would mean $r_{spike}^{2}\propto E_{o}^{2}$.

Spike times are generally calculated neglecting the cooling due to
evaporation. However, this cooling {\it is} significant when only a small
hot region is created, and the penetration depth has roughly the same size
as the lateral range of the ion, i.e. up to 10~--~20 keV in Xe $\rightarrow $
Au. On the other hand, for 100 keV Xe $\rightarrow $ Au, the lateral size of
the cascade is $\sim 60$ \AA , while the ion range is $\sim 135$ \AA\ (see
Table \ref{table:individual}). The flow of atoms to the surface starts at
about 0.1 ps, even though the ``ballistic'' part is not finished. The ion is
colliding further down from the surface, where the spike has not developed
yet, but liquid flow already started above this region.

As seen in Table \ref{table:averages}, Eq. \ref{tspike} understimates the spike 
lifetime in the bulk, because $R_c$ is used as the characteristic length scale. 
However, the agreement between Eq. \ref{tspike} and the MD results for the 
lifetime of surface spikes in cratering events is good, while the surface 
spike lifetime in non-cratering events is shorter, especially at higher 
energies. The outflow times from Eq. \ref{tflow0} are close to the spike 
times from Eq. \ref{tspike} for energies above 5 keV, making possible the 
crater formation by melt flow.

\subsection{Cratering probability}

\label{craterprob}

The cratering probability for a projectile with energy $E_{o}$, $P\left(
E_{o}\right) $, can be estimated as follows.

\[
P\left( E_{o}\right) =\int_{R_{i}}P\left( E_{o},R_{i}\right) dR_{i} 
\]

where the probability of creating a crater of radius $R_{i}$ is written as $%
P\left( E_{o},R_{i}\right) $. In order to simplify the analysis, we assume
that the probability of producing a crater with certain radius $R_{i}$ is a
delta distribution for $R_{i}=R_{c}\approx R_{l}$, $P\left(
E_{o},R_{i}\right) =P\left( E_{o},R_{c}\right) \delta \left(
R_{i}-R_{c}\right) $. This is especially valid at low energies, where
fluctuations in crater size are not as large as for higher energy, as was
also noted in the experiment of Merkle and J\"{a}ger \cite{Jag88}. This
probability can itself be split in three contributions. $P\left[ \varepsilon
\left( E_{o}\right) >\varepsilon _{c}\right] $ is the probability of
reaching the threshold energy density $\varepsilon _{th}$ in the subsurface
layer. $P\left[ t_{spike}>t_{flow}\right] $ is the probability of the spike
lifetime being longer than the outflow time. Finally, $P_{split}\left(
E_{o},R_{l}\right) \, $is the probability of the cascade splitting into two
cascades with each having not enough critical energy, or going too deep into
the sample. Then,

\begin{eqnarray}
P\left( E_{o}\right) &=& P\left( E_{o},R_{l}\right) =  \label{Pcratering} \\
&&P\left[ \varepsilon \left( E_{o}\right) >\varepsilon _{th}\right] \left[
1-P_{split}\left( E_{o},R_{l}\right) \right] P\left[
t_{spike}>t_{flow}\right]  \nonumber
\end{eqnarray}

Note that at low values of $E_{o}$, the threshold energy density should be $%
\varepsilon _{th}\sim 0.3\varepsilon _{U_{0}}=0.3 nU_{0}$, and that the term $%
P\left[ t_{spike}>t_{flow}\right] $ will be roughly 1. This is because the 
crater is created mainly by ``ballistic'' events, not melt flow.On the other 
hand, at high energy density, the critical energy is much lower, 
$\varepsilon_{c}\sim 3k_{B}T_{m}\sim 0.07\varepsilon _{U_{0}}$, and spike 
times are important. From the MD simulations, we find that $P_{split}\left(
E_{o},R_{l}\right)$ depends on energy. It is near 0 at low energies, and
increases to 0.2~--~0.4 at higher energies. Of course, for energies much
higher than here, it will eventually approach 1.

We show calculated values of $P\left( E_{o}\right)$ in Table \ref
{table:averages}. The values in the table are calculated by using 
$P\left[ t_{spike}>t_{flow}\right] = 1$. 
$P\left[\varepsilon \left( E_{o}\right) >\varepsilon _{th}\right] $ 
and $P_{split}$ were now obtained from MD, but could be obtained from 
a much simpler BCA calculation. The values for the calculated cratering 
probability compare well with the values from MD.

For $E_o > 20$ keV liquid flow is the main contribution and the energy
density threshold for crater formation is lower, leading to an enhanced
cratering probability. The cratering probability decreases rapidly for
$E_o > 50$ keV due to cascades being deep below the surface, and
splitting due to fast recoils. Our simple model reproduces all these
features, and extrapolating our results to $E_o=200$ keV gives
$P(200$ keV$)=0.3 \pm 0.1$ and $P(400$ keV)$=0.07 \pm 0.05$, which compares well 
with the experimental value of $P(400$ keV)$=0.03$ 
from Ref. \onlinecite{Bir2000}.
 
\section{Comparison to experiment}

\label{expt_comp}

The crater radius from our MD simulations can be seen in Fig. \ref{crmean}
as a function of bombarding energy. The radius can be well approximated by a 
$E_{o}^{1/3}$ dependence at low $E_{o}$, but at large $E_{o}$ there is a
saturation, which is related to saturation of the energy deposition in the
target. This energy deposition can be related to the known values for the
stopping power, $dE/dx$, and it is included in the figure. There are some
models that predict a dependence of the crater radius with $\left(
dE/dx\right) ^{1/2}$ (Refs. \onlinecite{Yama82,Bit87}) and this is the
dependence shown in Fig. \ref{crmean} as $R_{c}=1.4$\thinspace $x^{1/2}\,l$,
where $l=n^{-1/3}$, and $x=\left( dE/dx\right) \left( l/U\right) $.

The data from Birtcher {\it et al.} \cite{Bir96,Don97} is also shown in Fig. 
\ref{crmean}. It is clear that within the uncertainties the experimental and
simulated data agree very well, giving good confidence in the validity of
our simulations.

In Ref. \onlinecite{Don97} Donnelly and Birtcher discuss a criterion
for crater formation based on the available energy density. They only look
at high-energy cascades where spike times are much longer than the outflow
time to form a large crater (with the mean size expected for those
energies). This corresponds to our high energy region.

The cratering efficiency is much smaller in the experiment (5\%) than in our
MD simulations ($\sim 50$ \%). There are two likely reasons for this:

\begin{enumerate}
\item  Even though crystal orientation was not perfectly known in the
experiment, the incidence angle $\theta $ was about 15$^{\circ }$, so there
has been a significant amount of channeling (see Table \ref{rangetable}, and
note that both the mean range and straggling are larger for $\theta
=15^{\circ }$ than for $\theta =25^{\circ }$ ), which decreases the energy
deposition close to the surface and therefore the probability of cratering.

\item  The cratering probability of 0.05 was found from the ratio of
cratering probability to crater destruction cross section. This
cross-section could be larger than estimated due to the large ion beam
fluxes and possible {\it enhanced} atom mobility induced by the electron
bombardment.
\end{enumerate}

Donnelly and Birtcher also assume that the faceting of the crater sides
occurs slowly by diffusion \cite{Don97}. However, our simulations show
(see Fig. 2) that the craters already are faceted directly in the collision
cascades due to the crystal structure.

\vbox{
\begin{figure}[tbh]
\includegraphics{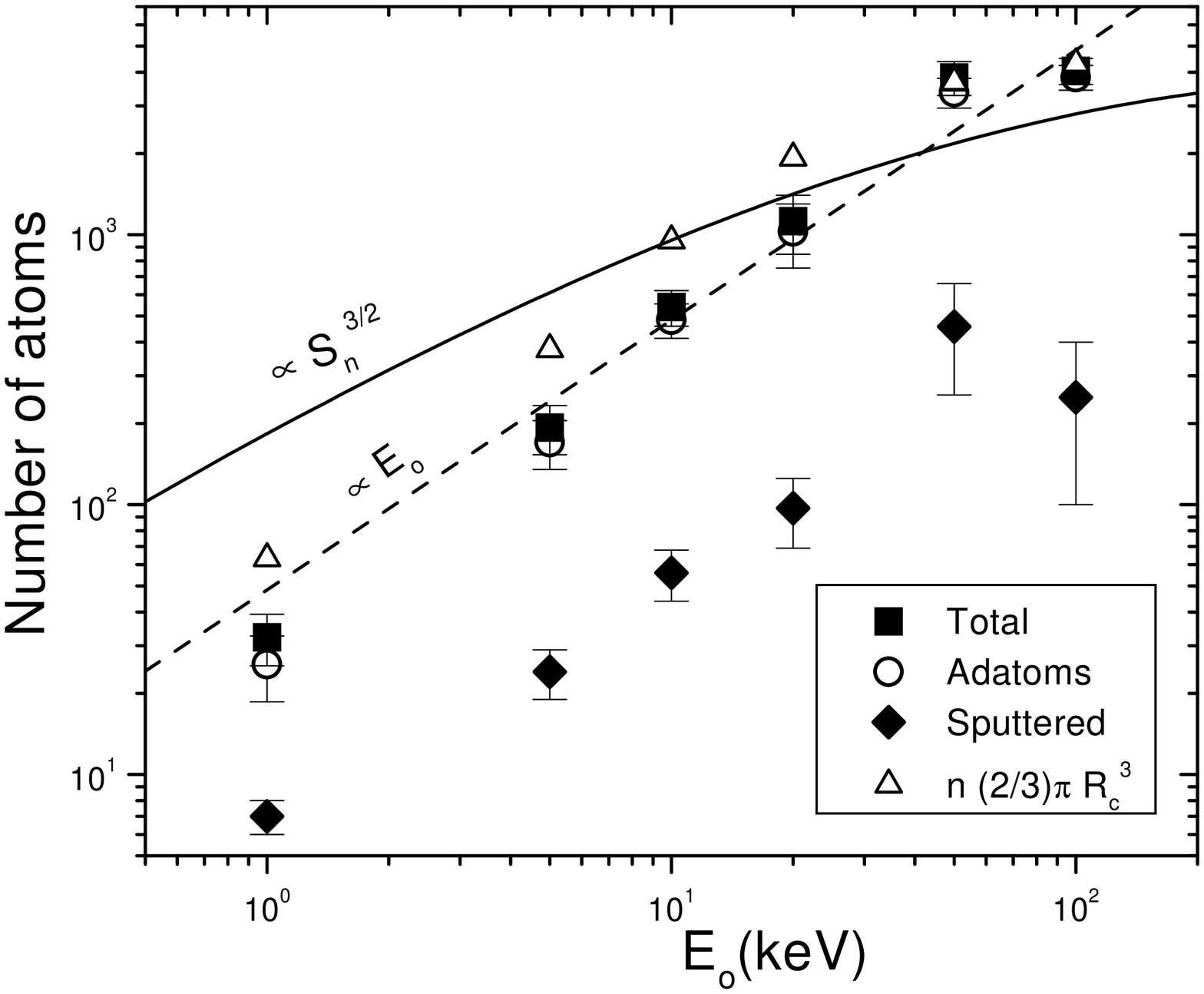}
\noindent a) \\
\nopagebreak \vskip 185pt minus 0pt \nopagebreak
\includegraphics{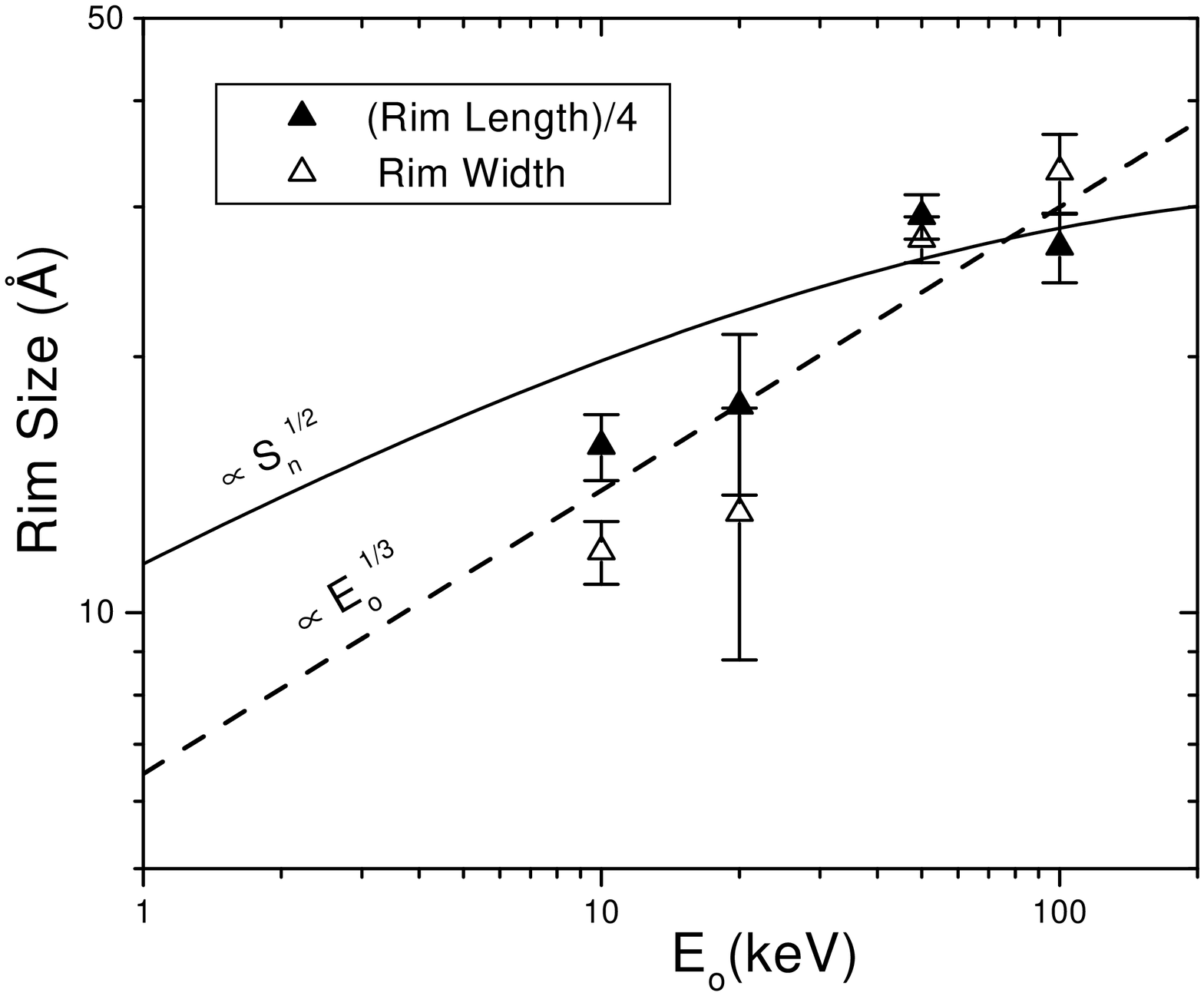}
\noindent b)\\
\nopagebreak \vskip 180pt minus 0pt \nopagebreak
\caption{ 
(a) Total number of atoms above the surface, adatoms, and sputtered
atoms as a function of the initial ion energy.
Note that because many of the sputtered
atoms sputter in hot clusters, and can subsequently evaporate from
the cluster and redeposit on the surface, this initial sputtering yield
does not exactly correspond to experimentally measured yields.
(b) Crater rim length and rim width versus the initial ion energy.
}
\label{rim-crater}
\end{figure}
}

The number of total atoms above the surface after bombardment and the size
of the crater rim can be seen in Figs. \ref{rim-crater} (a) and (b). As
expected from the behaviour of the crater radius, the number of atoms above
the surface is linear with $E_{o}$ at low energies and saturates at higher
energies

In the low energy regime we can cast our results in the following form:

\begin{equation}
\left( \frac{2\pi }{3}\right) nR_{cr}^{3}=A%
{\displaystyle {E_{o} \over U_{0}^{2}}}%
\label{rcratAu}
\end{equation}

\begin{equation}
N_{cr}=B\left( \frac{2\pi }{3}\right) n%
{\displaystyle {E_{o} \over U_{0}^{2}}}%
\label{ncrAu}
\end{equation}

Expressing energy in eV and density in \AA$^{-3}$, $A=\left( 1.39\pm
0.12\right) $ eV, $B=\left( 6.0\pm 0.75\right) $ eV\AA $^{3}$. Notice that $%
B\left( 2\pi /3\right) n=0.74\,$eV $<A$. The higher value of $A$ indicates
that the crater depth is typically smaller than the crater radius, as can be
seen in Fig.~\ref{rim-crater}~(a), where the open triangles represent the
left side of Eq. \ref{rcratAu}.

\vbox{
\begin{figure}[tbh]
\includegraphics{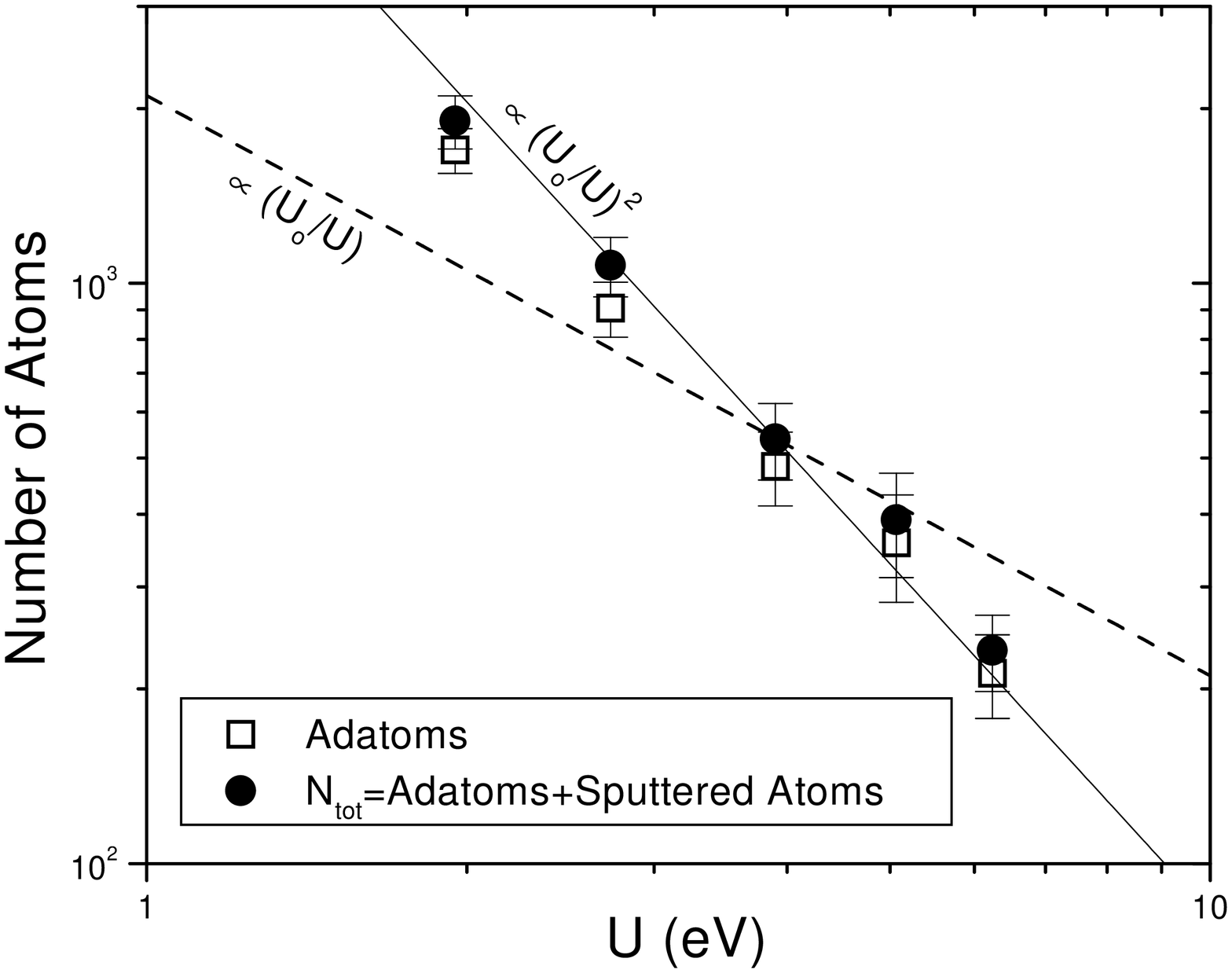}
\noindent a) \\
\nopagebreak \vskip 185pt minus 0pt \nopagebreak
\includegraphics{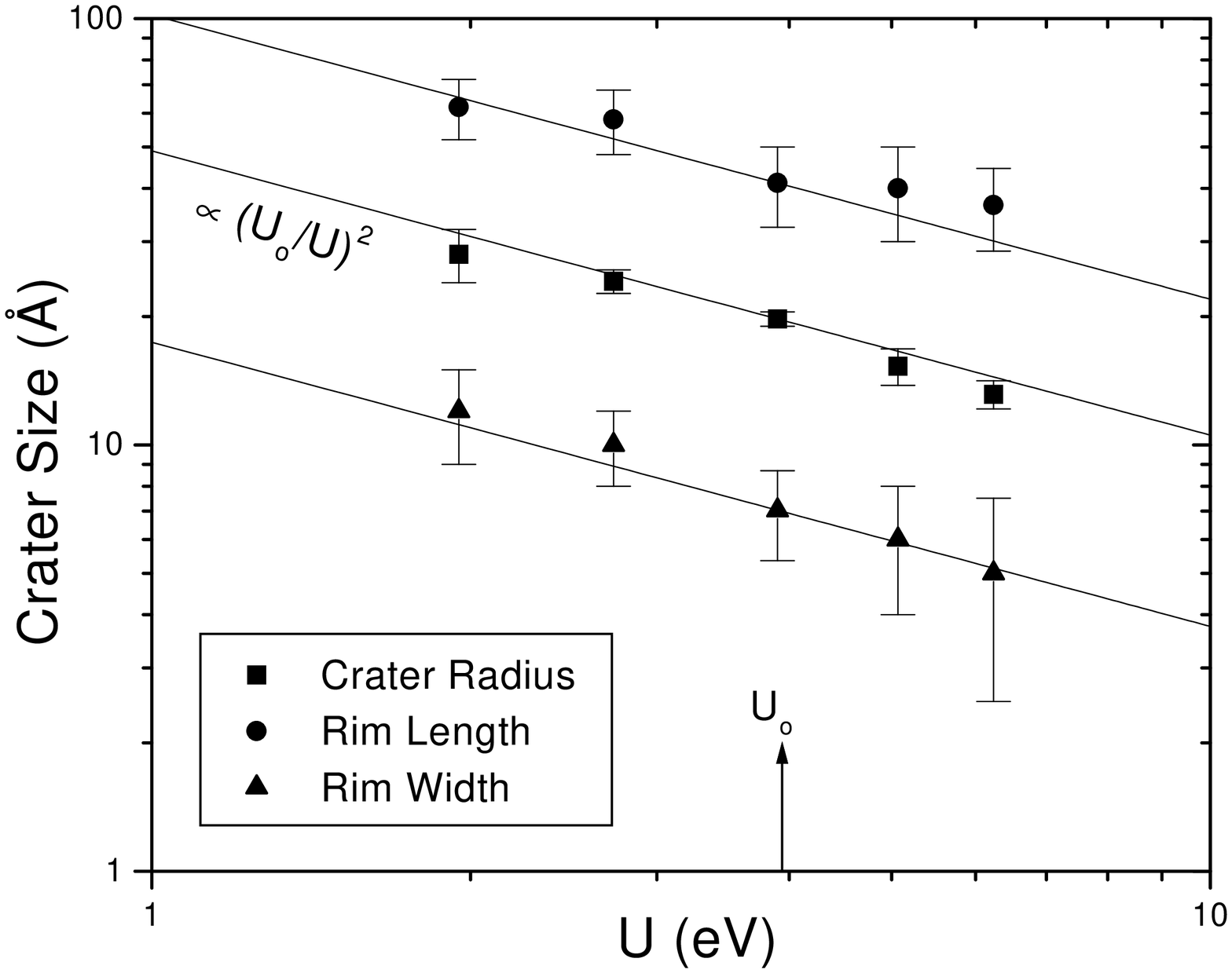}
\noindent b)\\
\nopagebreak \vskip 180pt minus 0pt \nopagebreak
\caption{
(a) Number of adatoms, and total number of atoms outside the surface, 
{\it vs.} binding energy $U$.
(b) Crater size {\it vs.} $U$. }
\label{size_binding}
\end{figure}
}

\section{Role of the binding energy of the target}

\label{binding_role}

Outside the ion beam community, it has generally been assumed, based on
scaling laws, that crater size scales as \cite{Gault75,insepovPRB} 
$U_{0}^{-1}$. This dependence was verified by macroscopic cratering 
events, like those of gas-gun experiments and astronomical objects where 
gravity can be neglected \cite{Gault75}. However, this has been recently 
challenged by results in both EAM Cu \cite{Ade00} and LJ\ solids 
\cite{Bri00a}, where the dependence was found to be $U_{0}^{-2}$. 
To test this, we repeated our 10 keV Xe bombardment simulations with 
the Au binding modified by changing the potential as explained in 
section II.

In Fig. \ref{size_binding} a) the number of atoms on top of the surface (as
explained in section \ref{MD-Analysis}) is plotted versus the binding
energy, and a quadratic dependence is found for bindings $U$ in the range 1.6%
$U_{0}$-0.5$U_{0}$. This quadratic dependence may result from a combination
of two factors. We found that the number of atoms in the melt {\it and} the
lifetime of the spike scale as $U^{-1}$. Therefore, if the total number of
adatoms scales as the product of the liquid atoms generated in the cascade
and spike lifetime, the result depends on $U$ as $U^{-2}$ (See also Footnote %
\onlinecite{tmeltnote}). In Fig. \ref{size_binding}-b the crater size is
plotted as a function of the binding energy, also showing a dependence of $%
U^{-2}$.

\section{Comparison with cluster bombardment}

\label{cluster_comp}

Since there are a number of experiments and simulations dealing with
cratering induced by cluster bombardment, we discuss some relevant cases.

For cluster bombardment at low $E_{c}$, the situation is such that the
energy is deposited in a roughly hemispherical region and the threshold for
crater formation is similar to what is found for atomic projectiles in the
high energy regime.


As mentioned in the introduction, Aderjan and Urbassek \cite{Ade00} recently
presented cratering results for Cu$_{n}\rightarrow $Cu where the number of
atoms in the crater was found to scale as 
\begin{equation}
N_{cr}=131E(keV)-656.  \label{ncrCu}
\end{equation}
In addition, a scaling with $U^{-2}$ was found. Using Eq. (\ref{ncrCu}) for
20 keV bombardment of Cu, gives $N_{cr}=1964$, while using the density and
binding of Cu in our Eq. (\ref{ncrAu}) gives $N_{cr}=1820\pm 230$, in an
excellent agreement with the previous estimate. Notice that the 20 keV Xe
bombardment deposits all its energy close to the surface, as the Cu$_{n}$
clusters do. The scaling found by Aderjan and Urbassek can not explain the
experimental results of Xe bombardment of Au, nor the results for
bombardment of C$_{60}$ on HOPG \cite{Aoki97_MRS}, since the crater size
increases always linearly with energy. This gives additional evidence for
the need of a more complex model. Besides, Aderjan and Urbassek suggested
that presence of viscosity may be the cause of the quadratic behavior with $U
$. However, for EAM liquids near the freezing point, the viscosity $\nu $
scales as $\nu \propto \sqrt{T_{m}}\propto \sqrt{U}$ 
(Ref. \onlinecite{March00}). If the
crater is formed mainly by outflow we can consider some simplified
cylindrical flow and use Poisson's equation, where the number $N$ of flowing
atoms is $N\propto \sqrt{\nu ^{-1}}\propto U^{-1/4}$. This would give only
an extra factor of 0.25 in the exponent and therefore it is quite unlikely
to be the reason for the transition from linear to quadratic behavior.

Typically, cluster bombardment simulations and experiments have been done at
low energy per bombarding atom. For C$_{60}$ bombardment of HOPG, the crater
``trace'' (corresponding roughly to the rim length in our simulations),
measured with an STM, was found to be proportional to $\left[ \left(
dE/dx\right) _{n}\right] ^{1/2}$, from 100 eV/atom up to 1 keV/atom, even
though the experimental errors are quite large \cite{Aoki97_MRS}. A
hemispherical crater is created, whose radius follows the law: $\left(
2/3\right) \pi R^{3}nE_{c}=E_{o}$, with $E_{c}\sim 0.05U_{0}$. This mean
energy is consistent with the low energy densities found in the molten
region.

All these results support our findings that spikes play a major role in
crater formation, and that a simple linear scaling of the crater volume with
the bombarding energy is not a good description except at very low energies.

\section{Comparison with models for other kinds of materials}

\label{othermater}

Even though most cratering studies have been performed in the regime where
the bombarding ion deposits most of its energy in elastic collisions, there
are a number of experiments \cite{papaleo,papaleo0} and simulations \cite
{david,david1,Bri99b} dealing with cratering in the regime where most of
the energy is initially deposited in the electrons of the target.
Experiments and simulations of cratering in solids made of large
biomolecules \cite{david,david1} suggested that the crater formation process
is mainly due to the large pressure pulse following electronic relaxation.
However, the mechanisms of surface erosion on condensed gas solids \cite
{Bri99b} seem to be controlled by thermal spikes, with pressure playing
only a secondary role in crater formation.

Condensed gas solids and other soft materials can be reasonably well
approximated by using simple 2-body potentials, as the Lennard-Jones
potential (LJ). The track of excitations can be modeled as a cylindrical
track of radius $r_{cyl}$, with atoms having some extra kinetic energy,
i.e., a cylindrical spike. For tracks of fixed radius, $r_{cyl}\sim 2l$,
with $l=n^{-1/3}$, there seems to be a critical energy density necessary for
crater formation, which is close to $nU$ \cite{Bri00a}. For rim formation,
the energy density needed is even higher, of the order of the bulk modulus
of the material or $\sim 9nU$. However, large craters can also be produced
at a low energy density, as in the case of 100 keV Xe$\rightarrow $ Au. In
LJ\ solids this has been seen in simulations of tracks, where $E_{c}\sim 0.8U
$ produces no crater for a spike radius $r_{spike}\sim 2l$, while for $%
r_{spike}\sim 5l$ the same energy density can produce large craters because
the spike lasts longer \cite{Bri99c}. LJ rare gases are comparatively
stiffer than metals, and the pressure pulse associated with the high
temperature spike takes a significant fraction of the energy when the energy
density is large, but it can be neglected for low energy densities. This
also {\it decreases the relative} lifetime of the spike as compared to
metals.

There are of course several differences to ion bombardment of metals, where
the spike radius will vary significantly with the energy of projectile, and
the energy deposition will not be uniform with depth, leading sometimes to
energy deposition profiles closer to spherical geometry. In track
simulations there is always a region at the surface, the top of the
cylindrical track, which is energized. On the other hand, in the case of ion
bombardment in the nuclear stopping regime, the peak in the energy
deposition occurs below the surface. However, in the regime with energy
above 10 keV for Xe bombardment, the created spike is initially roughly
cylindrical, and one would hope certain scaling still to be valid.

In EAM fcc metals the critical energy density needed for crater formation
for relatively narrow cascades is also close to $nU$, while for more
extended cascades much lower energy densities also produce cratering.
However, the energy density equivalent to the bulk modulus of Au is 1.04
eV/\AA {}$^{3}=4.54nU_{0}$. This energy density is never reached in the Xe
bombardment. On the other hand, the shear modulus is equivalent to $%
0.77nU_{0}$.

In the LJ track simulation, significant cascade splitting does not occur
because of the initial conditions and can be neglected. In addition, because
of the many body contribution in the EAM potential, the binding of small
clusters is larger than in the LJ\ potential, and this may lead to an
enhanced liquid flow to the surface.

In Fig. \ref{crater-LJ-EAM} we compare the current results on cratering by
Xe in EAM Au with those of cratering in a LJ solid by high-energy ions 
\cite{Bri00a}. To make the two cases comparable, we give the abscissa in terms
of $(dE/dx)(E_{0})(n^{-1/3}/U$ for an incident ion with energy $E_{0}$, and
the crater radius scaled by the characteristic length scale of the material $%
n^{-1/3}$. The figure shows that there is remarkably good agreement between
the scaling of the crater radius $R_{c}$ with the stopping power. The rim
width shown in the inset, however, does not follow any simple scaling law.
The likely reason is the different heat spike geometry (cylindrical {\it vs.}
hemispherical) leading to different crater shapes.

\vbox{
\begin{figure}[tbh]
\includegraphics{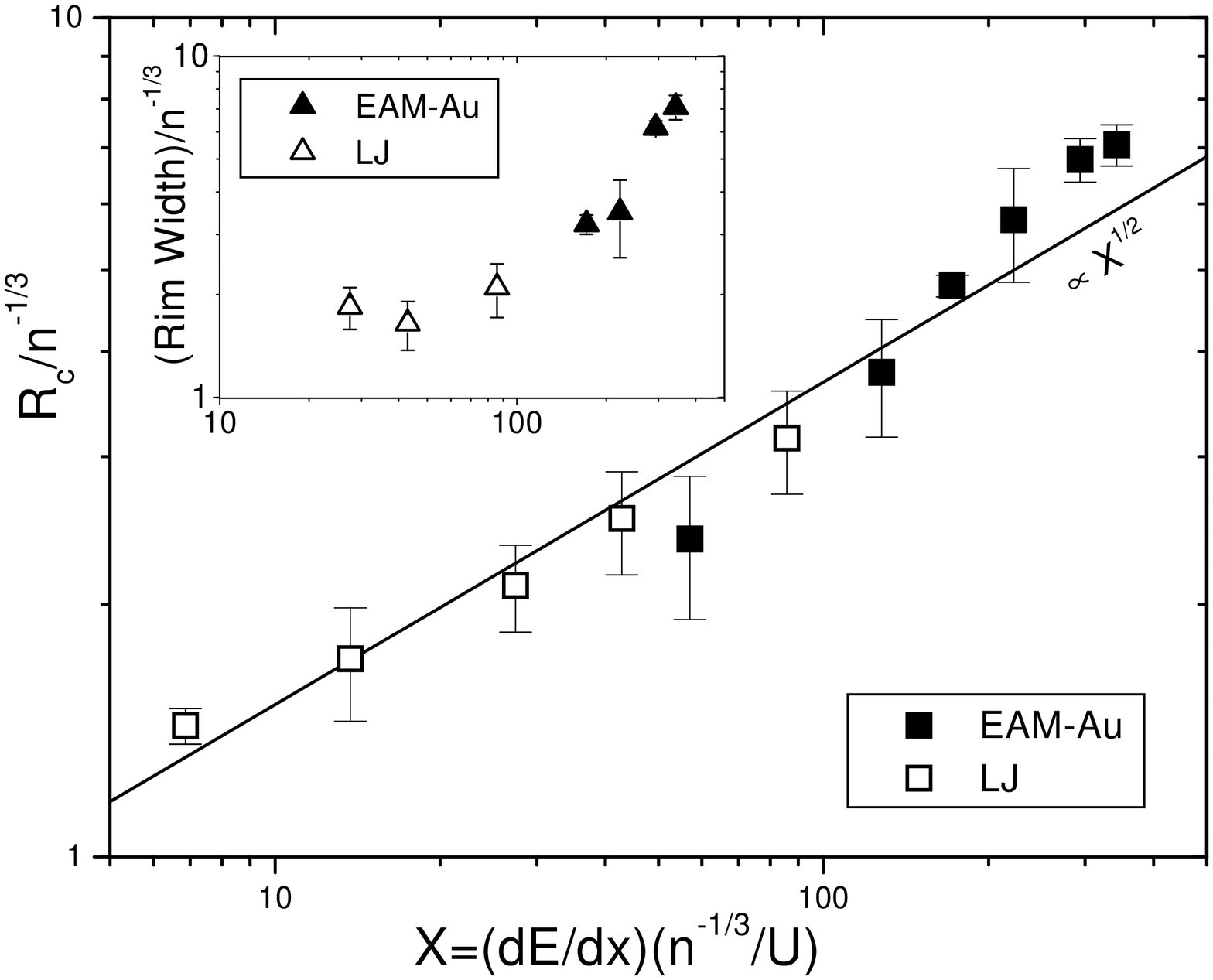}
\nopagebreak \vskip 210pt minus 0pt \nopagebreak
\caption{Crater dimensions versus scaled $dE/dx$ for 
both EAM Au and a LJ solid. The LJ results are from 
Ref. \protect\onlinecite{Bri00a}
}
\label{crater-LJ-EAM}
\end{figure}
}

Thus we see that despite of the major differences between metals and some
insulators which can be modeled as LJ solids, the crater sizes can still be
understood in the same framework, when the incident energy deposited into
atomic collisions (either directly or via electronic excitations) is
considered along with the stopping power of the material.

\section{Conclusions}

In conclusion, we have examined crater production by 0.4~--~100 keV Xe ions
impacting on Au using classical MD simulations. On the basis of our results,
we could show that the cratering mechanisms can not be understood in terms
of a single parameter (such as a single energy density), but rather in terms
of two energy regimes. In the low-energy regime (ion energies of $\sim 1 -
20 $ keV in the present system) the cascades which produce craters need to
have relatively high initial energy densities, $\sim 0.5 nU_0$. However, at
higher energies ($\gtrsim 100$ keV) the liquid formed in the heat spike can
become so long-lived that plenty of time is available for liquid flow, and a
much lower initial energy density ($\sim 0.2 nU_0$) suffices for crater
formation. These observations explains some apparent discrepancies between
previously presented cascade models.

We further demonstrated by direct simulation the importance of cascades
splitting into subcascades on the cratering probability, and presented an
analytical framework which can be used as a basis for further model
development accounting for cascade splitting, energy density thresholds, and
spike lifetimes. Furthermore, we showed that by using scaling laws and
parameters calibrated from MD, simple BCA codes such as SRIM can be used to
estimate cascade formation probabilities

We obtained excellent agreement with crater sizes measured experimentally in
the same system.

Comparison with simulations of cluster bombardment and crater formation in
LJ solids, which serve as models for some organic materials, showed that a
wide range of microscopic cratering events can be understood in the same
framework provided that appropriate scaling of the energy deposition and
length scales are used.

We also showed that macroscopic cratering laws behave quite differently from
the microscopic ones because of the importance of the liquid flow in the
microscopic system, but not in the macroscopic systems. Conversely, we note
that the results on cascade cratering may imply the need for a re-evaluation
of macroscopic scaling laws for materials which expand strongly on melting,
if the projectile can create a high enough energy density.

\section*{Acknowledgements}

We thank Prof. R. S. Averback, Prof. R. E. Johnson, and Prof. S. E. Donnelly
and Dr. R. C. Birtcher for useful discussions. Ms. L. Uusipaikka is
acknowledged for performing the 400 eV simulations. The research was
supported by the Academy of Finland under project No. 73722 and the Vilho,
Yrj\"{o} and Kalle V\"{a}is\"{a}l\"{a} foundation. 
Generous grants of computer time from the Center for Scientific Computing in
Espoo, Finland are gratefully acknowledged.

\bibliographystyle{prstyx}


\begin{thebibliography}{10}

\bibitem{Jag88}
W. {J\"ager} and K.~L. Merkle, Phil. Mag. A {\bf 57},  479  (1988).

\bibitem{papaleo}
R.~M. Papal\'{e}o, L.~S. Farenzena, M.~A. {de Araujo}, R.~P. Livi, M.
  Alurralde, and G. Bermudez, Nucl. Instr. Meth. Phys. Res. B {\bf 148},  126
  (2000).

\bibitem{papaleo0}
R.~M. Papal\'{e}o, Nucl. Instr. Meth. Phys. Res. B {\bf 131},  121  (1997).

\bibitem{balanzat}
F. Thibaudau, J. Cousty, E. Balanzat, and S. Bouffard, Phys. Rev. Lett. {\bf
  67},  1582  (1991).

\bibitem{bumps}
Q. Yang, T. Li, B.~V. King, and R.~J. MacDonald, Phys. Rev. B {\bf 53},  3032
  (1996).

\bibitem{Ade00}
R. Aderjan and H.~M. Urbassek, Nucl. Instr. Meth. Phys. Res. B {\bf 164-165},
  697  (2000).

\bibitem{Reimann-crat}
C.~T. Reimann, Nucl. Instr. Meth. Phys. Res. B {\bf 95},  181  (1994).

\bibitem{reimann-crater}
D.~A. Kolesnikov, C.~T. Reimann, and I.~V. Vorobyova, Nucl. Instr. Meth. Phys.
  Res. B {\bf 122},  255  (1997).

\bibitem{vorobyova}
I.~V. Vrobyova, Nucl. Instr. Meth. Phys. Res. B {\bf 146},  379  (1998).

\bibitem{Bir96}
R.~C. Birtcher and S.~E. Donnelly, Phys. Rev. Lett. {\bf 77},  4374  (1996).

\bibitem{Nor98c}
K. Nordlund, J. Keinonen\markthis, M. Ghaly, and R.~S. Averback, Nature {\bf
  398},  49  (1999).

\bibitem{Kyu99}
K. Kyuno, D.~C. Cahill, R.~S. Averback, J. Tarus, and K. Nordlund, Phys. Rev.
  Lett. {\bf 83},  4788  (1999).

\bibitem{Gha97}
M. Ghaly, K. Nordlund, and R.~S. Averback, Phil. Mag. A {\bf 79},  795  (1999).

\bibitem{Gha94}
M. Ghaly and R.~S. Averback, Phys. Rev. Lett. {\bf 72},  364  (1994).

\bibitem{aoki-thesis}
T. Aoki, Ph.D. thesis, Kyoto University, 2000,
  http://nishiki.kuee.kyoto-u.ac.jp/\symbol{126}t-aoki/papers/doctor/index.htm%
l.

\bibitem{Ave98}
R.~S. Averback and T. {Diaz de la Rubia},  in {\em Solid State Physics}, edited
  by H. Ehrenfest and F. Spaepen (Academic Press, New York, 1998), Vol.~51,
  pp.\ 281--402.

\bibitem{andersen}
H.~H. Andersen, A. Brunelle, S. Della-Negra, J. Depauw, D.~J. D, Y. {Le Beyec},
  J. Chaumont, and H. Bernas, Phys. Rev. Lett. {\bf 80},  5433  (1998).

\bibitem{yamada1}
I. Yamada, Mat. Chem. and Phys. {\bf 54},  5  (1998).

\bibitem{yamada2}
I. Yamada, J. Matsuo, N. Toyoda, T. Aoki, E. Jones, and Z. Insepov, Mat. Sci.
  and Engnr. A {\bf 253},  249  (1998).

\bibitem{yamada3}
I. Yamada, Nucl. Instr. and Meth. B {\bf 148},  1  (1999).

\bibitem{Colla2000}
T. Colla, R. Aderjan, R. Kissel, and H. Urbassek, Phys. Rev. B {\bf 62},  8487
  (2000).

\bibitem{Bri98}
E.~M. Bringa and R.~E. Johnson, Nucl. Instr. Meth. Phys. Res. B {\bf 143},  513
   (1998).

\bibitem{shulga-clusters}
V.~I. Shulga and P. Sigmund, Nucl. Instr. and Meth. B {\bf 47},  236  (1990).

\bibitem{averback-clusters}
R.~S. Averback, T.~D. de~la Rubia, H. Hsieh, and R. Benedek, Nucl. Instr. and
  Meth. B {\bf 59/60},  709  (1991).

\bibitem{averback-clusters1}
M. Ghaly and R.~S. Averback, {\em Mat. Res. Soc. Symp. Proc.} (Materials
  Research Society, Pittsburgh, 1992), Vol.~279, p.\ 17.

\bibitem{insepov-nato}
Z. Insepov and B. Kabdiev, {\em Intern. Conf. on Phys. and Chem. of Finite
  Systems: from Clusters to Crystals} (PUBLISHER, ADDRESS, 1992), Vol.~1, p.\
  429.

\bibitem{insepov}
Z. Insepov and I. Yamada, Nucl. Instr. and Meth. B {\bf 112},  16  (1996).

\bibitem{insepov1}
T. Aoki, J. Matsuo, Z. Insepov, and I. Yamada, Nucl. Instr. and Meth. B {\bf
  121},  49  (1997).

\bibitem{insepovPRB}
Z. Insepov, R. Manory, J. Matsuo, and I. Yamada, Phys. Rev. B {\bf 61},  8744
  (2000).

\bibitem{Nor98}
K. Nordlund, L. Wei\markthis, Y. Zhong, and R.~S. Averback, Phys. Rev. B (Rapid
  Comm.) {\bf 57},  13965  (1998).

\bibitem{Gault75}
D. Gault, J.~E. Guest, J.~B. Murray, D. Dzurisin, and M. Malin, J. Geophys.
  Res. {\bf 80},  2444  (1975).

\bibitem{lampson}
C.~W. Lampson, {\em Effects of Impact and Explosion} (Office of Scientific
  Research and Development, ADDRESS, 1946), Vol.~1, p.\ 110.

\bibitem{quinones98}
S. Quinones and L. Murr, Phys. Stat. Sol. A {\bf 166},  763  (1998).

\bibitem{explosion-cratering}
 in {\em Impact and explosion cratering}, edited by D.~J. Roddy and O. Pepin
  (Pergamon Press, New York, 1977).

\bibitem{Don97}
S.~E. Donnelly and R.~C. Birtcher, Phys. Rev. B {\bf 56},  13599  (1997).

\bibitem{Bir2000}
R. Birtcher, S. Donnelly, and S. Schlutig, Phys. Rev. Lett. {\bf 85},  4968
  (2000).

\bibitem{Nor97f}
K. Nordlund, M. Ghaly, R.~S. Averback, M. Caturla, T. {Diaz de la Rubia}, and
  J. Tarus, Phys. Rev. B {\bf 57},  7556  (1998).

\bibitem{Nature}
K. Nordlund, J. Keinonen, M. Ghaly, and R.~S. Averback, Nature {\bf 398},  49
  (1999).

\bibitem{And79}
H.~H. Andersen, Appl. Phys. {\bf 18},  131  (1979).

\bibitem{Foi86}
S.~M. Foiles, M.~I. Baskes, and M.~S. Daw, Phys. Rev. B {\bf 33},  7983
  (1986).

\bibitem{ZBL}
J.~F. Ziegler, J.~P. Biersack, and U. Littmark, {\em The Stopping and Range of
  Ions in Matter} (Pergamon, {New York}, 1985).

\bibitem{Daw93}
M.~S. Daw, S.~M. Foiles, and M.~I. Baskes, Mat. Sci. Rep. {\bf 9},  251
  (1993).

\bibitem{Lin10}
F.~A. Lindemann, Phys. Z. {\bf 11},  609  (1910).

\bibitem{Nor94b}
K. Nordlund, Comput. Mater. Sci. {\bf 3},  448  (1995).

\bibitem{SRIM2000}
J.~F. Ziegler, 2000, sRIM-2000.39 computer code, private communication.

\bibitem{Ave94b}
R.~S. Averback and M. Ghaly, J. Appl. Phys. {\bf 76},  3908  (1994).

\bibitem{Tho79}
D.~A. Thompson and S.~S. Johar, Appl. Phys. Lett. {\bf 34},  342  (1979).

\bibitem{oldbarriernote}
This estimate neglects the energy barrier needed for the outflow, which is
  expected to be much lower than the binding energy. If the barrier, $E_{b}$,
  is known, then $t_{outf}$ should be multiplied by
  $exp\left[+E_{b}/\left(\frac{2}{3}\overline{E}\right)\right]$.

\bibitem{noelphonon}
Because the most recent evidence indicates that energy transfer to the
  electronic subsystem occurs over timescales longer than those active in
  cascades, the electronic heat conductivity can be neglected at least as a
  first approximation. See Ref. \protect\cite{Nor98,Stu99}.

\bibitem{Yama82}
Y. Yamamura, Nucl. Instr. Meth. {\bf 194},  515  (1982).

\bibitem{Bit87}
I. Bitensky and E. Parillis, Nucl. Instr. Meth. Phys. Res. B {\bf 21},  26
  (1987).

\bibitem{Bri00a}
E.~M. Bringa, R. Papaleo, and R.~E. Johnson,   (2001), submitted.

\bibitem{tmeltnote}
This argument relies on the observation that the melting point scales linearly
  with the inverse of the cohesive energy, as it does in the simple
  modification of the potential used here. However, we have earlier found that
  by modifying only the repulsive part of the interatomic potential, the
  melting point can be modified with no change in the cohesive energy
  \protect\cite{Nor98}. Although it would be of some interest to examine the
  crater size scaling for such a change, this is not directly instructive for
  the main point of the present paper, namely the comparison with macroscopic
  scaling laws, and hence will not be done here.

\bibitem{Aoki97_MRS}
T. Seki, T. Aoki, M. Tanomura, J. Matsuo, and I. Yamada, Materials Chemistry \&
  Physics {\bf 54},  143  (1997).

\bibitem{March00}
N.~H. March, Phil. Mag. A {\bf 80},  1335  (2000).

\bibitem{david}
D. Fenyö and R.~E. Johnson, Phys. Rev. B {\bf 46},  5090  (1992).

\bibitem{david1}
R.~E. Johnson, B.~U. Sundqvist, A. Hedin, and D. Fenyö, Phys. Rev. B {\bf 40},
  49  (1989).

\bibitem{Bri99b}
E.~M. Bringa and R.~E. Johnson, Nucl. Instr. Meth. Phys. Res. B {\bf 152},  267
   (1999).

\bibitem{Bri99c}
E.~M. Bringa, R.~E. Johnson, and M. Jakas, Phys. Rev. B {\bf 60},  15107
  (1999).

\bibitem{Stu99}
A.~E. Stuchbery and E. Bezakova, Phys. Rev. Lett. {\bf 82},  3637  (1999).

\end{thebibliography}


\end{document}